\documentclass[12pt]{article}
\usepackage{graphicx}

\def\napoli{Logunov Institute for High Energy Physics, NRC Kurchatov Inst., Protvino, RF}

\def\Title#1{\begin{center} {\Large #1 } \end{center}}
\def\Author#1{\begin{center}{ \sc #1} \end{center}}
\def\Address#1{\begin{center}{ \it #1} \end{center}}

\newenvironment{Abstract}{\begin{quotation}  }{\end{quotation}}
\newenvironment{Presented}{\begin{quotation} \bigskip 
      \begin{center}\begin{large}}{\end{large}\end{center} \end{quotation}}
\def\Acknowledgements{\bigskip  \bigskip \begin{center} \begin{large}
             \bf ACKNOWLEDGEMENTS \end{large}\end{center}}

\begin{document}
\begin{titlepage}

\vfill
\Title{Towards Precision Measurement 
of
Elastic Scattering at U-70 }
\vfill
\Author{ Sergey P. Denisov and \underline{Vladimir A. Petrov} }
\Address{\napoli}
\vfill
\begin{Abstract}
A short version of the recent proposal to measure elastic scattering of protons at the Protvino proton synchrocyclotron U-70.
\end{Abstract}
\vfill
\begin{Presented}
Presented at EDS Blois 2017, Prague, Czech Republic, June 26-30,2017
\end{Presented}
\vfill
\end{titlepage}
\def\thefootnote{\fnsymbol{footnote}}
\setcounter{footnote}{0}

\section{Introduction}

Let me first remind you of the 70-GeV proton synchrotron U-70 at the Logunov IHEP in Protvino. Launched in operation in the fall of 1967 as the most powerful accelerator at that time, it produced some outstanding results among which I only recall the discovery of the growth of $ \sigma_{tot}^{K^{+}p} $ , scaling in hadronic inclusive cross-sections and observation of the anti-tritium production.
Last measurements of the elastic scattering were made near 1975 \cite{ant} . Fig.1 shows the differential cross-sections and deviations from the reference  exponential form $ \exp(Bt+Ct^{2}) $. Since then such deviations were observed on repeated occasions
at various machines and energies \cite{sel} .

\begin{figure}[hbt!]
	\begin{center} 
		\includegraphics[width=0.98\textwidth, height=7.5cm]{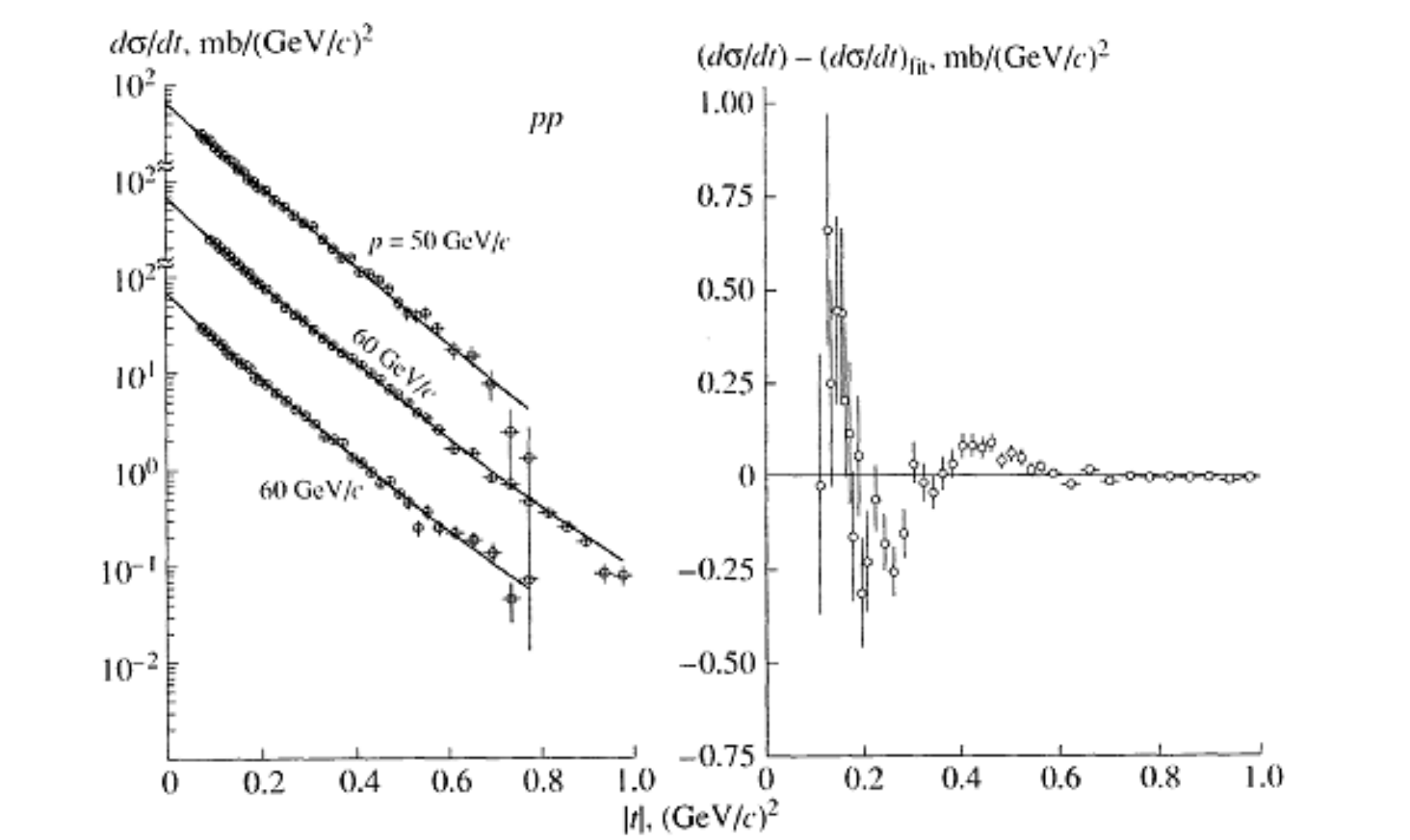}
		\caption{\label{fig:Fig1} }
	\end{center} 
\end{figure}  

In Fig.1 are shown differential cross-sections from \cite{ant} fitted with a squared exponential (left) and the difference between  the 60 GeV data and their fitted values (right)

 Most recent example is the TOTEM result \cite{tot}. Suggested physical reasons of such deviations extend from the influence of a close $ \pi\pi $ threshold in t-channel and up to a residual colour forces of the Van der Waals type. There were also opinions that such oscillations are  but  statistical fluctuations having no profound physical implications.
To clarify this dilemma we have decided to repeat measurements with much greater accuracy which is now allowed by new, much improved experimental means.
Modern state-of-art methods for particle detection, identification and data acquisition and analysis make it possible to perform , at the IHEP accelerator,
studies of elastic $pp$ scattering at the lab energy 50 GeV on the basis of a record statistical data sample embracing about $ 10^{9} $ events at a high momentum-transfer resolution. This would give us an opportunity to test the earlier results 
\cite{ant} and also to provide measurements in the still unexplored region $ -t > 1 GeV ^{2} $. The choice of energy 50 GeV is motivated by the fact that last years the accelerator operated at this maximum energy. One could ask why to measure effects at 50 GeV when much higher energies (LHC) are available? At least two reasons:first, due to the analyticity in energy, strong interaction amplitudes 
correlate even at large difference in energy values; second,the U-70 energies ( $ O(10 GeV) $ are exactly in the region where the nucleon valence cores just begin to detach from each other. At the LHC they are still not very far from each other.
So, from \textit{this }perspective the difference is not so significant.
\section{Layout of the experiment}

The setup of the proposed experiment is pictured in Figure2.
\begin{figure}[hbt!]
	\begin{center} 
		\includegraphics[width=0.98\textwidth, height=4cm]{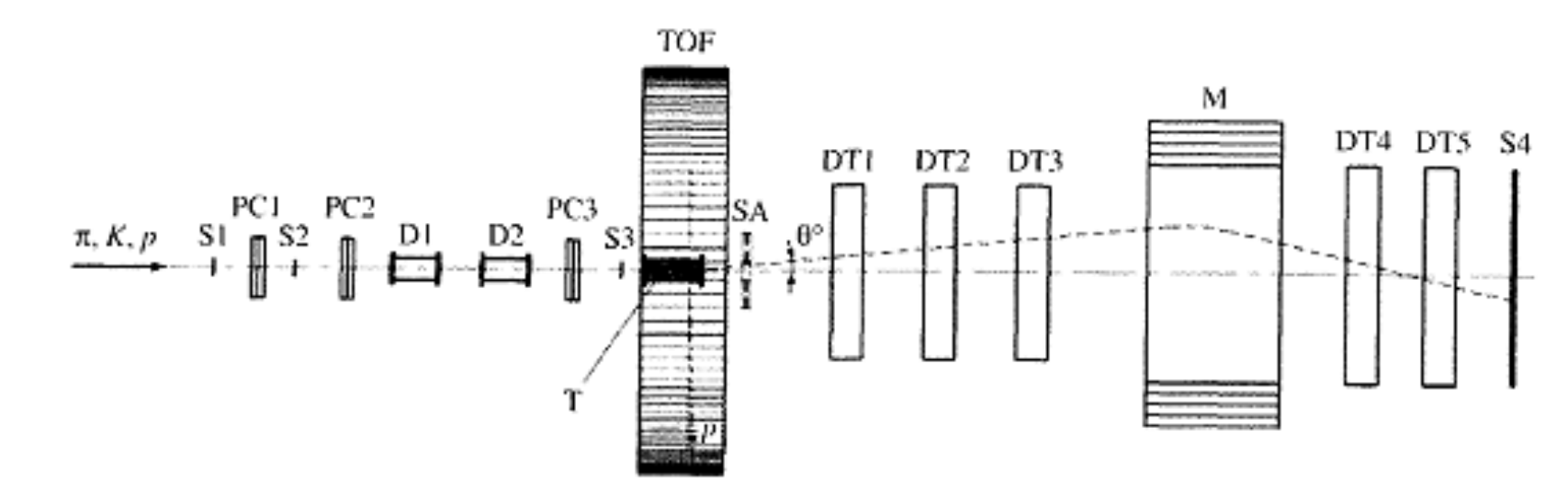}
		\caption{\label{fig2:layout} Layout of the proposed experimental setup.}
	\end{center} 
\end{figure}  

At the figure of layout we designated: scintillation counters(S1-S4), anticoincidence counters (SA), proportional chambers (PC1-PC3),differential Cherenkov counters (D1\&D2), drift-tube stations (DT1-DT5), scintillation counters for measurements of the recoil proton time of flight(TOF), spectrometric magnet(M)
A target (T) is manufactured of Mylar and is filled with gaseous hydrogen at a pressure of up to 20 atmospheres.The useful target length is 1.5 m.At the gas pressure of 20 atm. and the diameter 100 mm the thickness of the target cylindrical wall is 1.0 mm of Mylar.At the mentioned pressure the target contains 0.27 $g/cm^{2}   $ of hydrogen or $ 1.6\cdot10^{26} $ protons/$ cm^{2} $.At the proton flux of $ 3\cdot 10^6  s^{-1}$ the number of elastic events is $ 3.7\cdot10 ^{3} $. With extention of the beam over 2 sec.per accelerator run(20 days) this yields $ 1.4\cdot10^{9} $ events (see Table 1 for the binning in t-intervals).

\begin{figure}[h!]
\centering

\includegraphics[width=0.6\textwidth]{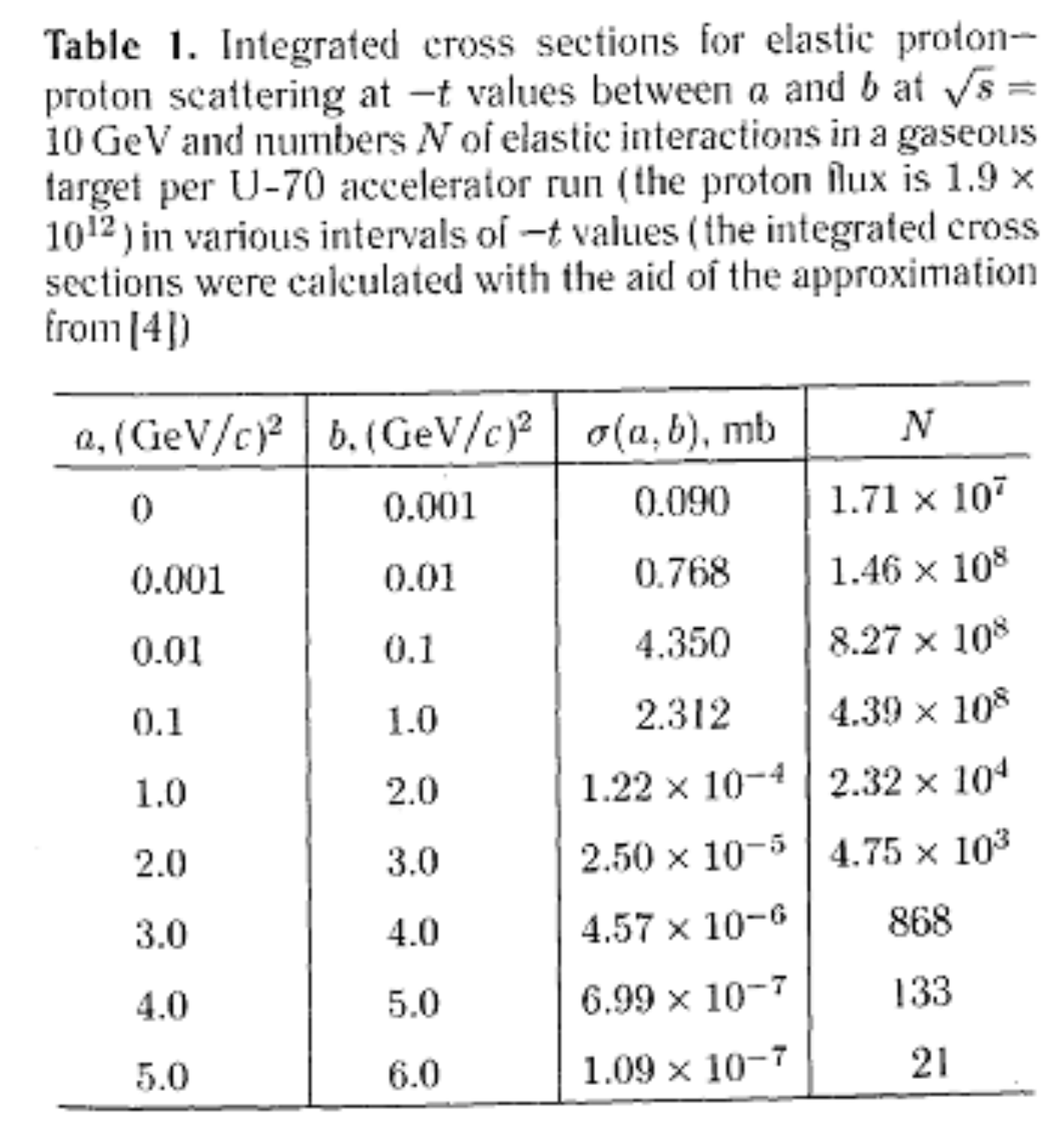}

\label{fig:errors}
\end{figure}
\begin{figure}[h!]
\centering
\includegraphics[width=0.7\textwidth]{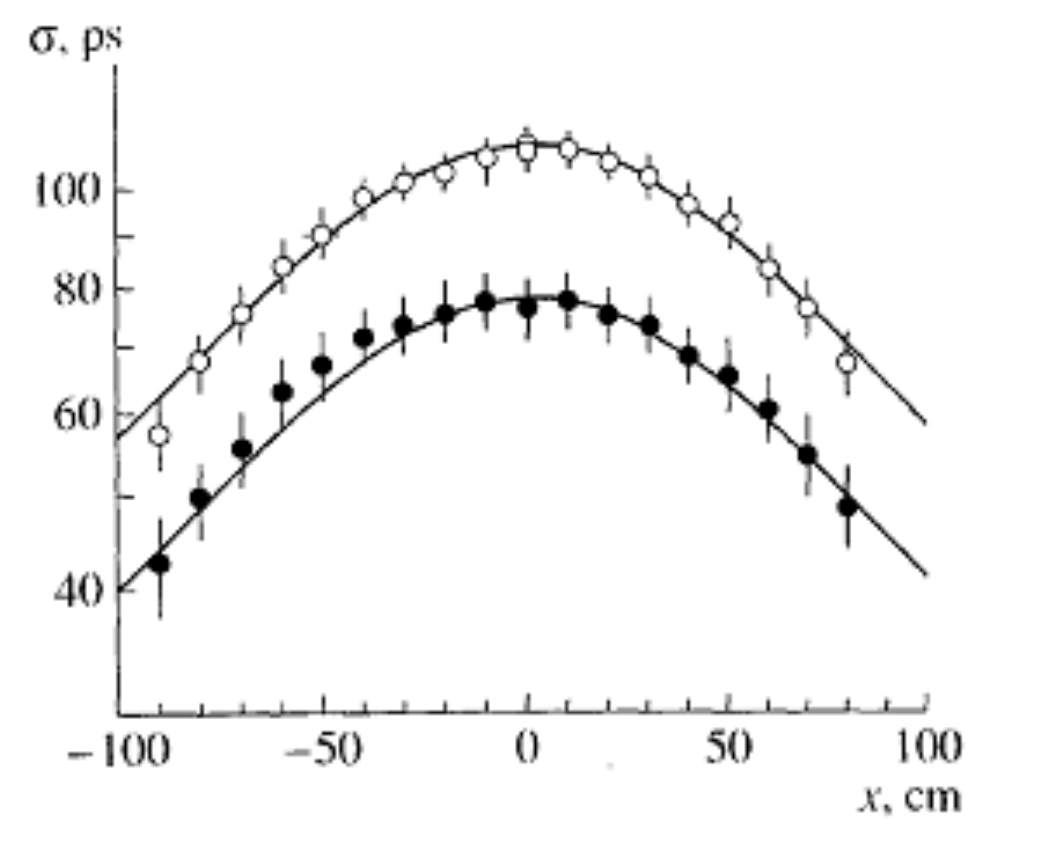}
\caption{Time resolution of one and two 2-m counters featuring HR2020 photomultiplier tubes vs the locus of particle propagation.}
\label{fig:table}
\end{figure}

\begin{figure}[h!]

\includegraphics[width=0.85\textwidth]{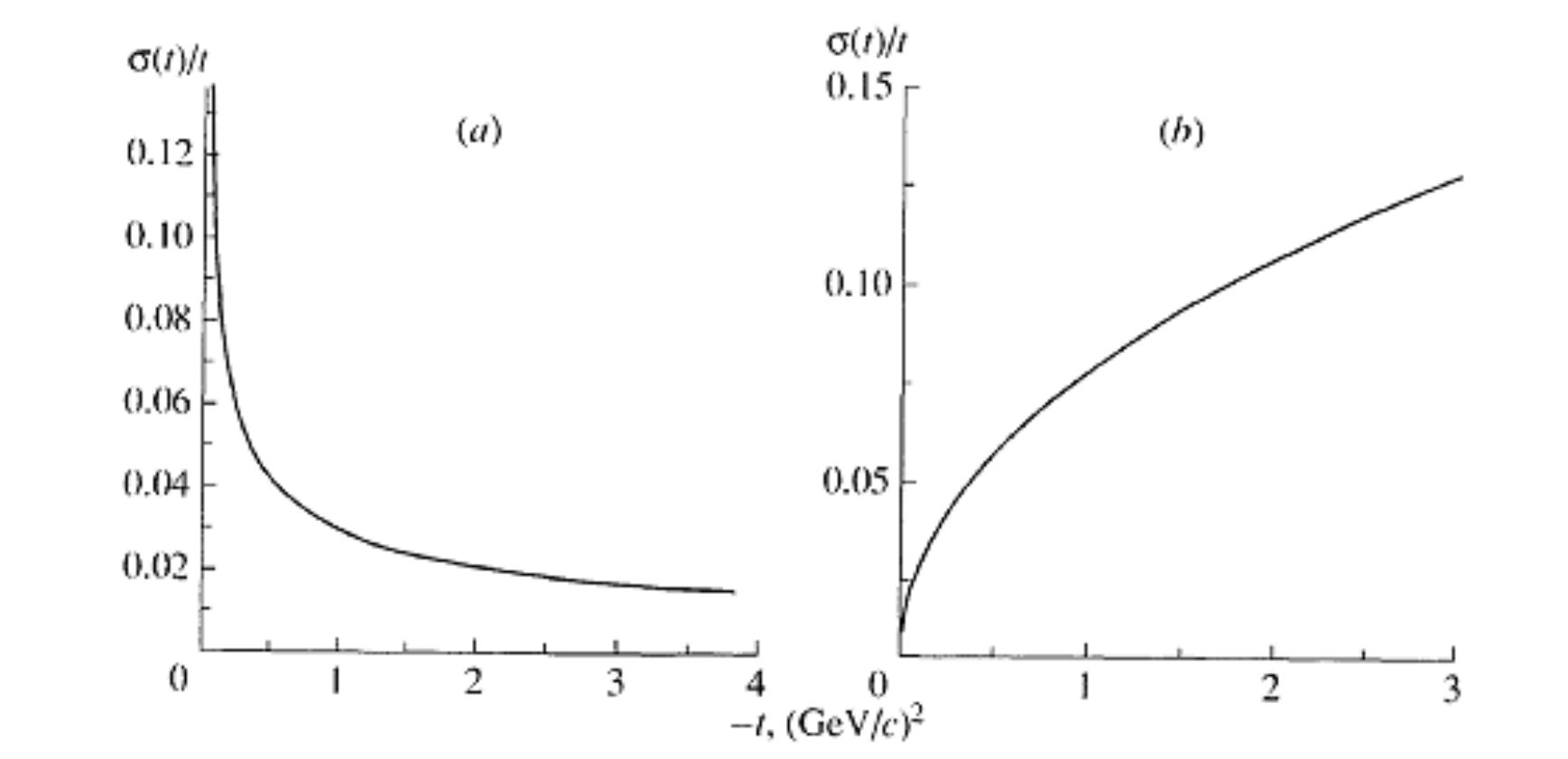}
\caption{Errors in determining $ -t $ }
\label{fig:table22}
\end{figure}
Near 60\% of the recoil protons should have enough energy to be detected in TOF.
With allowance for the setup efficiency the number of detected elastic events 
amounts to $ 8\cdot10^{8} $ with more than $ 10^{5} $ in the region $ -t>1GeV^{2}. $
The target is surrounded by a TOF system formed of 128 scintillation counters
(time resolution $\approx 100 ps  $ )that are arranged at a distance of 1.2 m 
from the target centre and which are intended for determining the speed of recoil protons by their time of flight and their energy by the ionization loss.The counter length is 2 m. Each counter ifs viewed from the two sides by R 1828-01 photomultiplier tubes.The amplitude and arrival time are measured for each multiplier-tube signal. A BC404A is used in the counters. The counter time resolution vs the locus of relativistic particle propagation is shown in Fig.3.

One can hope for a better time resolution in the case of employing an R1828-01 photomultiplier tube whose signal has a steeper front edge.

Estimates of the errors in determining  $ -t $ by the scattering angle of a $ 50GeV/c $ proton are shown in Fig.4a. Fig. 4b shows resolution from the determining momentum transfers by the recoil proton TOF.
\section{Conclusions}

All the necessary methodic work and all detector investigation, and also full MC simulation with use of GEANT4 are completed.

• There are the system of the proton abortion with help of the bent crystal, beam channel and spectroscopic magnet.
•  Drift tubes and relevant electronics are being produced.
•  There is scintillator for all counters, but only a half of photomultipliers  (HAMAMTSU).
•  A half of the necessary amount of  electronics (CAEN) for amplitude ant time analysis of signals from photomultipliers are purchased.
•  Production of the HV system for PMTs is in progress.

We need a 1.5 year and some 200 kUSD for completion of the facility.
For the studies in the CN region we have got practically everything and physical run can start in a year.

More detailed description of the proposal can be found in \cite{den}

\Acknowledgements
We are greatful to the organizers of the EDS 17 for invitation to give this talk.

\end{document}